\documentclass[reprint,twocolumn]{revtex4-1}%
\usepackage{graphicx}
\usepackage{bm}
\usepackage{latexsym}
\usepackage{epsf}
\usepackage{rotating}
\usepackage{epsfig,graphics,rotate,color}
\usepackage{wrapfig}
\usepackage{amssymb}
\usepackage{amsmath}
\usepackage{amsfonts}
\usepackage[T1]{fontenc}
\usepackage{accents}
\newcommand*{\bbar}[1]{\accentset{\tiny{(}-\tiny{)}}{#1}}

\begin{document}


\title{A Sterile Neutrino Search with Kaon Decay-at-rest}
\author{J. Spitz}
\affiliation{Massachusetts Institute of Technology, Cambridge, MA 02139, USA}

\begin{abstract} 
Monoenergetic muon neutrinos (235.5~MeV) from positive kaon decay-at-rest are considered as a source for an electron neutrino appearance search. In combination with a liquid argon time projection chamber based detector, such a source could provide discovery-level sensitivity to the neutrino oscillation parameter space ($\Delta m^2\sim 1~\mathrm{eV}^2$) indicative of a sterile neutrino. Current and future intense $\gtrsim$3~GeV kinetic energy proton facilities around the world can be employed for this experimental concept.
\end{abstract}
\maketitle

\section{Introduction}
A set of experiments and analyses hints towards the existence of at least one sterile neutrino. Most notably, the LSND~\cite{lsnd} collaboration has reported a 3.8$\sigma$ excess of $\overline{\nu}_e$ events from a $\overline{\nu}_\mu$ decay-at-rest source. The measurement is indicative of neutrino oscillations around $\Delta m^2\sim 1~\mathrm{eV}^2$, inconsistent with the solar/reactor~\cite{SuperKsolar, SNO, Kamland} and atmospheric/accelerator~\cite{SuperKatmos, SoudanII, k2kosc, MinosCC1,*MinosCC2,*MinosCC3,*MinosCC4,*MinosCC5} oscillation results and the three neutrino picture in general. A ``sterile'' neutrino, which does not couple to the Z-boson~\cite{zwidth}, has been introduced as a possible explanation for the anomalous oscillation signal. 

The LSND result is not alone. The MiniBooNE experiment has recently probed the $\Delta m^2\sim 1~\mathrm{eV}^2$ parameter space with neutrinos and anti-neutrinos. The accelerator-based MiniBooNE experiment searched for $\nu_\mu$ to $\nu_e$ oscillations and excluded the two neutrino oscillation hypothesis in the LSND allowed region at the 98\% confidence level~\cite{miniosc}. However, MiniBooNE has also probed the parameter space with anti-neutrinos and finds $\overline{\nu}_e$ appearance consistent with two neutrino oscillation in the $\Delta m^2 \in 0.1-1.0~\mathrm{eV}^2$ range. The excess in the $475<E_{\nu}<1250~$MeV energy range has a 0.5\% probability of being due to background and is consistent with LSND~\cite{nubarminiosc}. 

The ``reactor anti-neutrino anomaly'' is another hint towards the possible existence of a sterile neutrino~\cite{Mueller,Lassere}. The predicted reactor $\overline{\nu}_e$ spectra, applicable to almost all reactor-based neutrino experiments, have been re-examined and the ratio of the observed rate to the new predicted rate is found to be below unity at the 98.6\% confidence level. The apparent disappearance of MeV-scale reactor anti-neutrinos is consistent with LSND and an oscillation signal near $\Delta m^2\sim 1~\mathrm{eV}^2$, although the result is for $\overline{\nu}_e$ disappearance ($\overline{\nu}_e \rightarrow \overline{\nu}_{\not{e}}$) rather than the related $\overline{\nu}_e \rightarrow \overline{\nu}_\mu$ appearance process.

Models with both one sterile neutrino (3~active~+ 1~sterile) and two sterile neutrinos (3+2) have been invoked to explain the anomalies~\cite{Sorel:2003hf, Karagiorgi1,*Karagiorgi2, Schwetz, Giunti}. Despite the aforementioned results, however, the existence of the sterile neutrino(s) is still an open question. A definitive measurement of the parameter space favored by LSND, MiniBooNE anti-neutrino, and the reactor anomaly is required~\cite{whitepaper}. An experimental idea to probe neutrino oscillations near $\Delta m^2\sim 1~\mathrm{eV}^2$ using a kaon decay-at-rest source of $\nu_\mu$ along with a large liquid argon time projection chamber (LArTPC) detector is presented.

\section{Experimental concept}
Charged kaons at rest produce a monoenergetic 235.5~MeV $\nu_\mu$ in 63.6\% of decays. A detector placed $\mathcal{O}$(100~m) from a stopped kaon source can search for LSND-like $\nu_e$ appearance events from $\nu_\mu$ (originating with the decay-at-rest $K^{+} \rightarrow \mu^+ \nu_{\mu}$) oscillation in a narrow reconstructed energy window around the expected monoenergetic signal. 

The signature of a $\nu_e$ appearance event is the charged current, quasi-elastic interaction $\nu_e n \rightarrow e^- p $ at 235.5~MeV. These signal events compete with the analogous $\nu_\mu$ interaction $\nu_\mu n \rightarrow \mu^- p$ as well as background charged current interactions of $\nu_e$ from the three body decays $K^{+} \rightarrow \pi^0 e^+ \nu_{e}$ (BR=5.1\%) and $K^{0}_L \rightarrow \pi^\pm e^\mp \bbar{\nu}_{e}$ (BR=40.6\%). Although such events can nominally be considered a background for the $\nu_e$ appearance search, they can be used for flux and cross section measurements beneficial to the oscillation analysis.


The Booster Neutrino Beamline (BNB) at Fermilab is a pulsed source of 8~GeV kinetic energy protons. The beamline nominally services the MiniBooNE~\cite{minidet} and SciBooNE~\cite{sciboonecc} detectors and will provide protons for MicroBooNE~\cite{microboone} beginning in 2014. These experiments rely on pion decay-in-flight neutrino production for cross section measurements and oscillation sensitivity. The BNB is currently providing approximately 3$\times 10^{20}$~protons on target (POT) per year. The beam delivers about 4$\times 10^{12}$~POT at 3-5~Hz with a total spill time window of 1.6~$\mu$s.

Kaons are produced via a large copper target block placed directly in front of the primary proton beam. Kaons and other unstable charged particles created in the primary proton-copper and secondary interactions come to rest quickly and subsequently decay. Copper has been chosen due to its high thermal conductivity as well as its low radiation length ($X_0=1.43$~cm), which significantly reduces intrinsic $\nu_e$ background from $K^{+}$ and $K^{0}_L$ decay-in-flight. There is no magnetic focusing horn required.

\begin{figure}[h!]
\begin{center}
\begin{tabular}{c c}
\hspace{-.7cm}
\includegraphics[scale=.27]{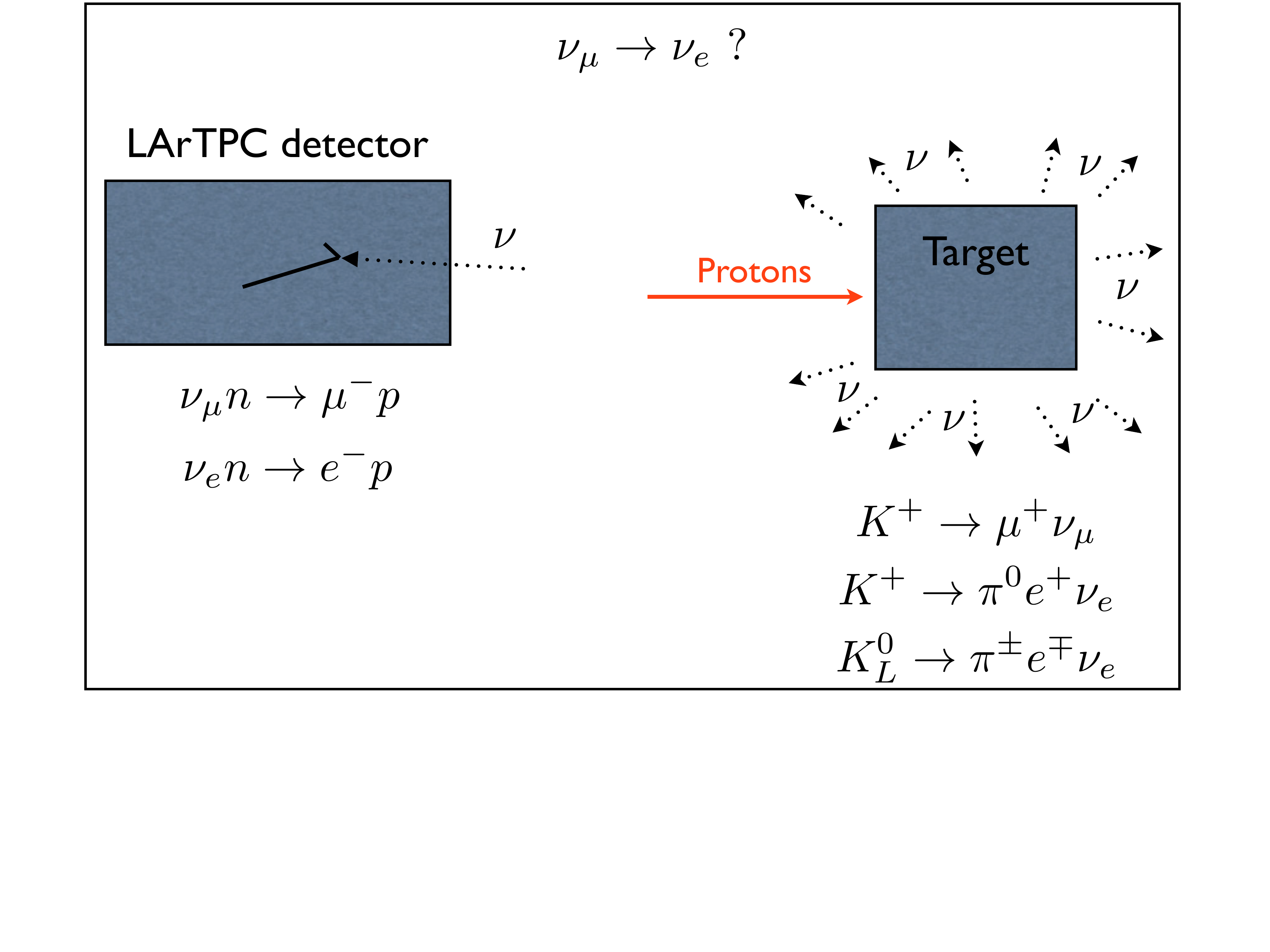}
\end{tabular}
\end{center}
\vspace{-2.4cm}
\caption{A schematic of the experimental design. The proton-on-target induced kaon/neutrino production location is seen on the right and the LArTPC-based neutrino detector is seen on the left. The most relevant kaon decay modes and dominant neutrino interactions are shown as well.}
\label{fig:schematic}
\end{figure}

A 2~kton LArTPC is considered as the neutrino detector for this study. The detector is envisioned 160~m from the target in a direction opposite or nearly opposite the primary beam. This orientation reduces decay-in-flight induced $\nu_e$ background. The baseline has been chosen in consideration of sensitivity to the LSND allowed region as well as event rate given the $1/r^2$ dependence of the neutrino flux. Note that a baseline of $\sim$240~m from the neutrino creation point corresponds to oscillation maximum for 235.5~MeV neutrinos at the LSND best-fit $\Delta m^2=1.2~\mathrm{eV}^2$. A schematic of the experimental layout is shown in Fig.~\ref{fig:schematic}.

The BNB can be considered an example beamline for the experimental concept described here. There are a number of other facilities around the world that can be used for such an experiment. The main requirements are that the proton beam have $\gtrsim$3~GeV kinetic energy and that it be high power. Both requirements translate into the single statement that an intense source of stopped kaons is needed for this design. An exposure of 2$\times 10^{21}$ protons on target is used for the oscillation sensitivity estimates here. Note that the total 8~GeV proton flux after the planned proton improvement plan at Fermilab will exceed 2$\times10^{17}$~protons per hour~\cite{protons}. The detector mass and beam exposure have been chosen in order to provide 5$\sigma$ sensitivity to most of the LSND allowed region. However, the numbers/parameters employed in this study are scalable to other beamlines, detectors, configurations, and experimental assumptions in general.

\section{Flux, cross section, and nuclear effects}
\begin{figure}[h!]
\begin{center}
\begin{tabular}{c c}
\includegraphics[scale=.47]{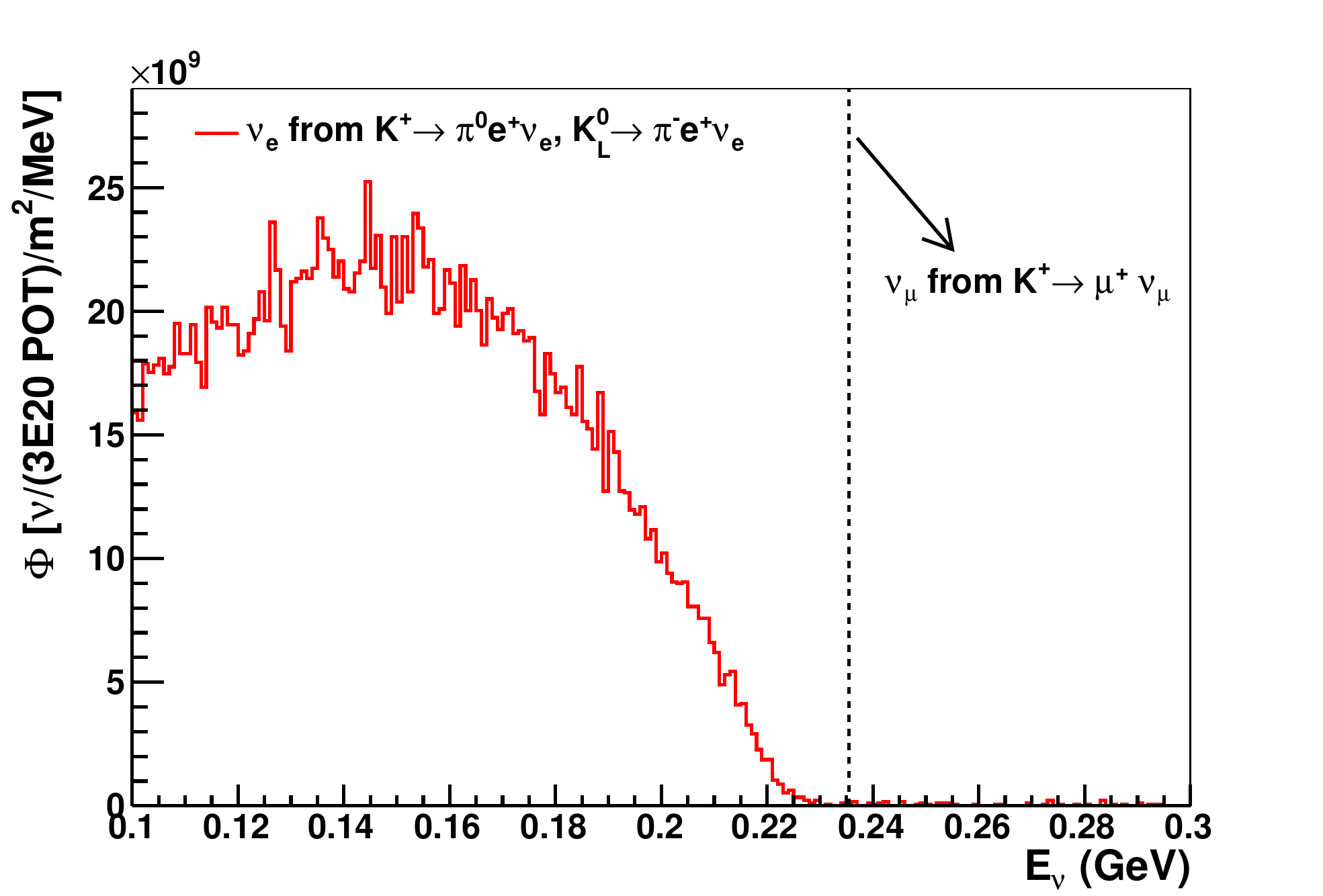}
\end{tabular}
\end{center}
\vspace{-.5cm}
\caption{The electron neutrino flux expected at 160~m. The dominant contribution is from charged kaon decay-at-rest. The monoenergetic $\nu_{\mu}$ flux (3.5$\times 10^{13}~\nu/(3\mathrm{E}20~\mathrm{POT})/\mathrm{m}^{2}$) location is also shown. }
\label{fig:flux}
\end{figure}

The 235.5~MeV $\nu_\mu$ flux and all-energy $\nu_e$ background is determined with the MARS~\cite{mars1,*mars2,*mars3,*mars4} and Geant4~\cite{geant4} (Quark-gluon string precompound with Bertini cascade model) simulation packages. Protons ($T_{\mathrm{p}}=$8~GeV) are directed onto a large pure copper target in the positive $z$ direction. Backwards going [$\cos(\theta_z)<-0.75$] neutrinos are considered as signal candidates for the purposes of event tabulation. The loose angular restriction is imposed in order to allow some detector placement flexibility--although a ``detector opposite beam direction'' policy is optimal. Target cooling and a detailed design of the target geometry have not been throughly examined in this study. The monoenergetic $\nu_{\mu}$/proton yield is 0.049 and 0.038, according to MARS and Geant4, respectively. The following study employs the Geant4 yield; the MARS result can be considered a reference value. The background flux prediction, alongside the  monoenergetic $\nu_\mu$ flux, is shown in Fig.~\ref{fig:flux}. It is useful to point out the similarity of this experimental idea to that of a neutrinoless double beta decay search: an excess of events is sought near the endpoint of a well predicted and measured background distribution.

A proper treatment of the nuclear physics affecting the initial neutrino interaction as well as the interaction products visible to the detector is vital for understanding the attainable neutrino oscillation sensitivity in this design. This experiment depends on the efficient identification and precise calorimetric reconstruction of $\nu_e$-induced charged particle tracks and the nuclear physics relevant to neutrino interactions plays a significant role in determining how well such events can be reconstructed.

Neutrino interactions on an argon target are simulated using the NuWro event generator~\cite{nuwro1,*nuwro2}. NuWro has been chosen due to its use of argon spectral functions in describing the momentum and binding energy distributions of nucleons within the argon nucleus~\cite{arspectral}. A spectral function implementation is known to be a more accurate description of the (argon) nucleus than one based on the relativistic Fermi gas model~\cite{benhar}. However, the related impulse approximation, in which (1) the neutrino interacts with a single nucleon and (2) the products are propagated through the nucleus, does not necessarily describe the interaction properly, especially at low energy transfer~\cite{arspectral}. The approximation is a general problem in the simulation of neutrino events; the spectral function model offers the best chance of the accurate simulation of neutrino-nucleus interactions at these energies. 

\begin{figure}[h!]
\begin{center}
\begin{tabular}{c c}
\includegraphics[scale=.47]{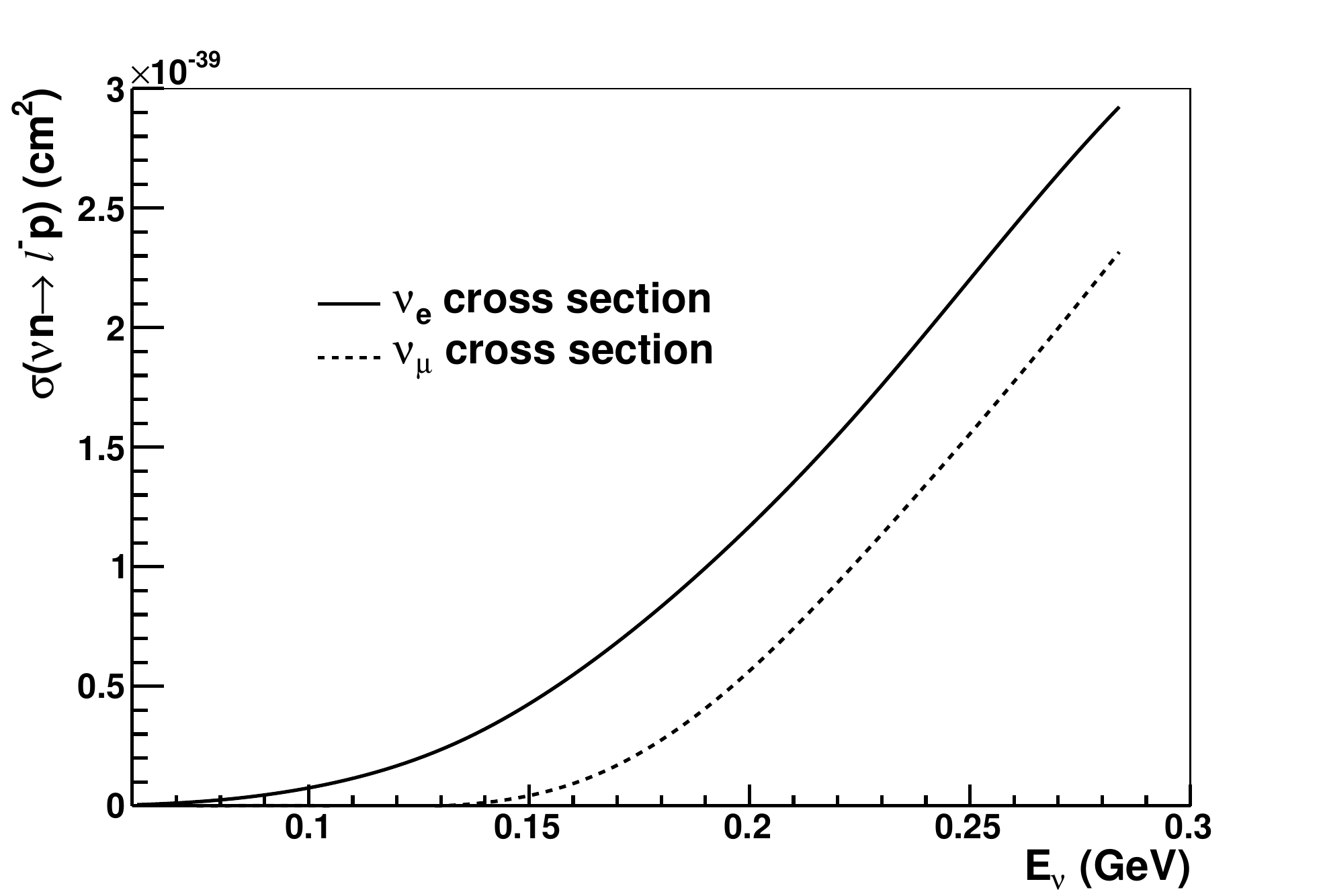}
\end{tabular}
\end{center}
\vspace{-.5cm}
\caption{Electron and muon neutrino cross sections from the NuWro neutrino event generator. The cross section is reported ``per target nucleon (neutron)''.}
\label{fig:xsec_nuwro}
\end{figure}

NuWro also simulates intra-nuclear interactions in which the hadron(s) produced in the neutrino interaction are allowed to reinteract with nucleons in the nucleus. Such reinteractions can result in multiple ejected nucleons/hadrons and ultimately impact neutrino energy reconstruction. A description of NuWro's intra-nuclear simulation can be found at Ref.~\cite{nuwro2}. The GENIE neutrino event generator~\cite{genie} has been used as a cross check of the nuclear model and the neutrino event simulation in general. The $\nu_\mu$ and $\nu_e$ charged current cross sections at 235.5~MeV employed in NuWro are $1.3\times10^{-43}$~m$^{2}$/neutron and $1.9\times10^{-43}$~m$^{2}$/neutron, respectively. The $\nu_e$ cross section in GENIE is about 25\% higher than the NuWro prediction. This is not unexpected, as the spectral function implementation generally reduces the (differential) cross section, most noticeably at low energy transfer~\cite{arspectral}. Note that the multi-nucleon component of quasi-elastic-like neutrino interactions at these energies~\cite{miniccqe,martini}, not simulated here, can act to enhance the cross section by more than 10\%. The NuWro cross sections across the relevant energy range are shown in Fig.~\ref{fig:xsec_nuwro}.

\section{Detector resolution}
The nominally distinctive monoenergetic $\nu_e$ oscillation signal is obscured by nuclear physics and detector resolution effects, both working to reduce the (available) precision in reconstructed neutrino energy. This $\nu_e$ appearance search requires the precise calorimetric reconstruction of the electron-induced electromagnetic shower and the associated outgoing proton or protons, recalling that multiple nucleons can be emitted due to intra-nuclear effects. The ICARUS collaboration estimates that the energy resolution for electromagnetic showers in a LArTPC~\cite{showerreco} from 50-5000~MeV can be parameterized with $\frac{\Delta E}{E}$=0.33/$\sqrt{E (\mathrm{MeV})}+0.012$. The parameterization is used here along with an all-outgoing-proton kinetic energy resolution ($\frac{\Delta T}{T}$) of 10\% for Gaussian smearing the simulated signal and background outgoing particles' energy, imitating reconstruction resolution. The proton resolution estimate is loosely based on an ICARUS experiment exposed to a neutrino beam and their measured value of $\frac{\Delta T}{T}=3.3\%$ for 50~MeV kinetic energy protons~\cite{wanfprd}. Neutrons are considered undetectable, lost energy and few-MeV-scale nuclear de-excitation gammas are not simulated in this study. A summary of the most relevant experimental assumptions and parameters is shown in Table~\ref{assumptions}.

\begin{table}[t]
\label{base_opti}
  \begin{center}
    {\footnotesize
      \begin{tabular}{|c|c|} \hline 
        Proton target  & Copper  \\  \hline
        $\frac{\nu_\mu (235.5~\mathrm{MeV})}{\mathrm{proton}}$ yield at $T_{\mathrm{p}}=$8~GeV  & 0.038 \\  \hline
        Exposure  & 2$\times 10^{21}$ protons on target  \\  \hline
        Baseline  & 160~m  \\  \hline
        Neutrino target  & $^{40}$Ar (22 neutrons)  \\  \hline
        Neutrino target mass  & 2~kton  \\  \hline
        Detection efficiency  & 100\%  \\  \hline
        $\nu_e~\sigma_{\mathrm{CC}}$ at 235.5~MeV  & $1.9\times10^{-43}$~m$^{2}$/neutron \\
        \hline
        $\nu_\mu~\sigma_{\mathrm{CC}}$ at 235.5~MeV  & $1.3\times10^{-43}$~m$^{2}$/neutron \\
        \hline
        $\frac{\Delta E}{E}$ for e$^{-}$ reconstruction & 0.33/$\sqrt{E (\mathrm{MeV})}+0.012$  \\ \hline
        $\frac{\Delta T}{T}$ for proton reconstruction & 0.10  \\ \hline
        Background syst. uncertainty  & 25\%  \\  \hline
      \end{tabular} 
      \caption{The most relevant parameters employed in this study.}\label{assumptions}
}
\end{center}
\end{table}

Asymmetric smearing effects due to nuclear physics and missing energy from outgoing neutrons created via intra-nuclear scattering renders the mean reconstructed neutrino energy ($E'_{\mathrm{rec}}=E_{\mathrm{electron}}+\sum_i^{n}T_{i,\mathrm{proton}}$, where $n$ is the number of protons) lower than the true neutrino energy ($E_{\mathrm{true}}$). Binding energy is the largest source of this lowering. The most probable value of the $E'_{\mathrm{rec}}$ distribution is made to coincide with $E_{\mathrm{true}}=235.5$~MeV with the addition of 45~MeV: $E_{\mathrm{rec}}=E'_{\mathrm{rec}}+$45~MeV. This correction constant, applied to all $\nu_e$ candidate events at all energies, is irrelevant to the oscillation sensitivity reported here--it is merely added so that reconstructed neutrino energy and true neutrino energy can be directly compared. Of course, the shapes of the signal and background $E_{\mathrm{rec}}$ distributions are critical for sensitivity to oscillations. The actual relationship between $E_{\mathrm{rec}}$ and $E_{\mathrm{true}}$, including all relevant detector and nuclear effects, can eventually be understood with the thousands of monoenergetic $\nu_\mu$ events and many hundreds of intrinsic $\nu_e$ events expected during data taking. 

\begin{figure}[h!]
\begin{center}
\begin{tabular}{c c}
\includegraphics[scale=.47]{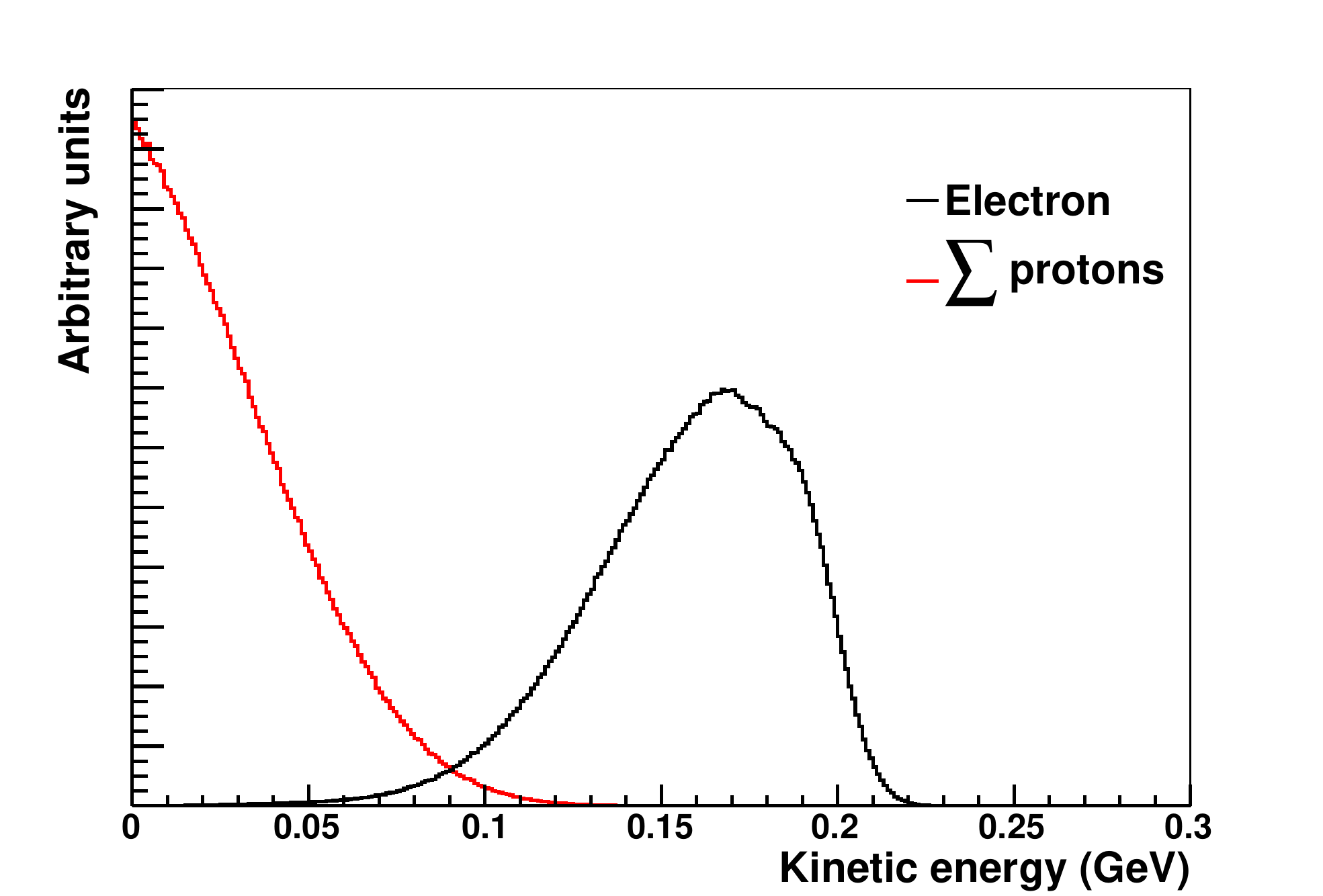}
\end{tabular}
\end{center}
\vspace{-.5cm}
\caption{The reconstructed kinetic energy of the outgoing electron and reconstructed total kinetic energy of all outgoing protons in 235.5~MeV electron neutrino events simulated with NuWro. Note that detector resolution and nuclear effects are included and events with no outgoing protons (due to proton absorption in the nucleus) do not enter the proton distribution.}
\label{fig:kinenergy}
\end{figure}

\begin{figure}[h!]
\begin{center}
\begin{tabular}{c c}
\includegraphics[scale=.47]{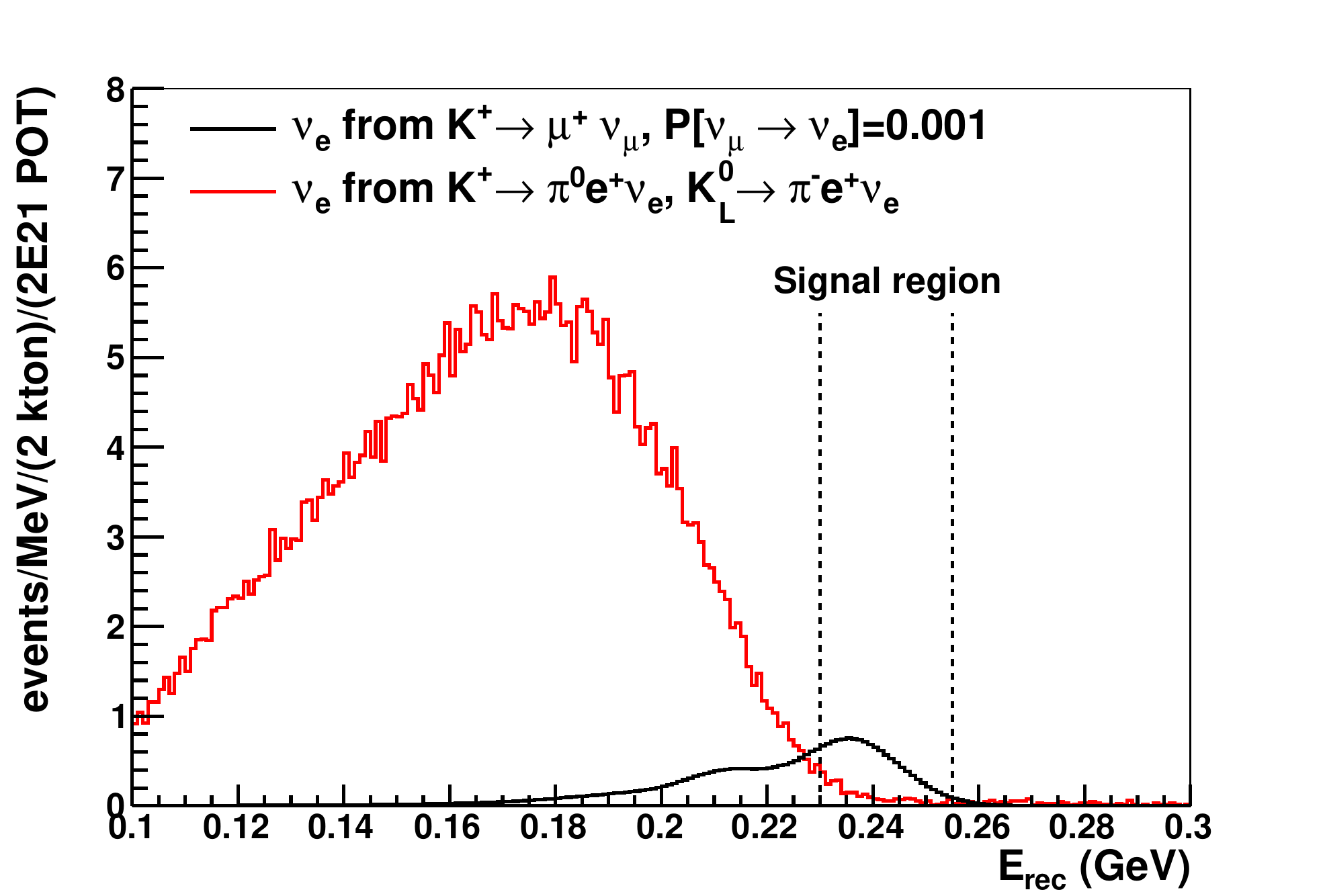}
\end{tabular}
\end{center}
\vspace{-.5cm}
\caption{Electron neutrino rate with detector resolution and nuclear effects included. The vertical lines designate the signal region.}
\label{fig:rate_nuwro}
\end{figure}

The kinetic energy of the outgoing particles in 235.5~MeV $\nu_e$ events is shown in Fig.~\ref{fig:kinenergy}. Assuming 100\% detection efficiency, the expected rate in terms of $E_{\mathrm{rec}}$, including smearing effects due to detector resolution and nuclear physics, can be seen in Fig.~\ref{fig:rate_nuwro}. The signal distribution's two hump shape comes from the nuclear shell structure and the neutron energy levels described by the argon-specific spectral function.

\section{Backgrounds}
The main contribution to the $\nu_{e}$ appearance background is from intrinsic $K^{+} \rightarrow \pi^0 e^+ \nu_{e}$ decay-at-rest. The maximum neutrino energy for this decay is 226~MeV, about 10~MeV lower than the monoenergetic signal. Contributions from decay-in-flight $K^{+}$ and $K^{0}_L \rightarrow \pi^\pm e^\mp \nu_{e}$ are also important and are apparent beyond the three body positive kaon decay's endpoint. The relevant $\nu_e$ background contributions to the flux can be seen in Fig.~\ref{fig:flux}. 

Although the uncertainty in kaon production is rather large, as exemplified by the $\sim$25\% discrepancy between the MARS and Geant4 predictions, background measurements are available $\textit{in-situ}$ when the experiment turns on. Background constraints come from the measured $K^{+}$ three body decay-at-rest $\nu_e$ spectrum and its well predicted shape as well as measurements of decay-in-flight $\nu_e$ background outside of the signal region. Charged kaon induced $\nu_e$ and monoenergetic $\nu_\mu$ events can also be used to determine the cross section ratio $\sigma_{\nu_{\mu}}/\sigma_{\nu_{e}}$ for precisely transforming measured rate into an oscillation probability.

Electron anti-neutrino events ($\overline{\nu}_e p \rightarrow e^+ n $), usually originating from $K^{0}_L \rightarrow \pi^+ e^- \overline{\nu}_{e}$, are a negligible background compared to the $\nu_{e}$ contribution due to (1) the difference in cross section and (2) the minimal $\overline{\nu}_e$ flux from $K^{-}$ production and subsequent three body decay.

A simple single bin, counting experiment is employed to discern a potential monoenergetic oscillated $\nu_e$ signal from background. All $\nu_e$-like events that enter the signal energy region of $E_{\mathrm{rec}}\in 230-255$~MeV are considered to have $E_{\mathrm{true}}=235.5$~MeV for the purposes of the oscillation analysis. Approximately 45\% of the true $\nu_e$ signal remains after employing this strict reconstructed energy requirement. The range has been chosen to provide sensitivity to the LSND allowed region with a single bin experiment, in consideration of maximizing signal-to-background and having enough events left over to discern a signal. Further optimization is possible as more knowledge becomes available. The actual experimental signal region can be refined with the thousands of expected non-signal reconstructed $\nu_\mu$ events in and near this energy range as well as $\sim$450~$\nu_e$ events between $E_{\mathrm{rec}}\in 100-230$~MeV. Along with signal region tuning, a more sophisticated analysis beyond the simplistic single bin, counting experiment can be performed for oscillation sensitivity improvement.

With a 2~kton detector, 2$\times 10^{21}$ POT exposure, and an oscillation probability of $P(\nu_\mu \rightarrow \nu_e)=0.001$, approximately 16~$\nu_e$-like (signal+background) events above a background of 3~events are expected. The signal events are searched for among an abundant set of distinctive monoenergetic $\nu_\mu$ events. The 13 true signal events in our example can be compared to approximately 9000 $\nu_\mu$ events expected in the same energy window. Obviously, the experimental concept described here relies heavily on the ability to differentiate the two classes of events. All future LArTPC-based electron neutrino appearance experiments rely heavily on the ability to differentiate charged current muon neutrino events from electron neutrinos ones. The differentiation is largely based on event topology with the muon and electron producing a characteristic track and shower, respectively, along with the decay and capture topology unique to the muon. Furthermore, calorimetric techniques can be used to differentiate muon-induced gammas, as in the case of a radiative muon capture, from electrons. Although a muon-electron misidentification with a LArTPC can occur in rare circumstances~\cite{microboone}, it is vanishingly rare for a misidentified true 235.5~MeV $\nu_\mu$ event to reconstruct as a $\nu_e$ event in the tight energy window required to be considered signal ($E_{\mathrm{rec}}\in 230-255$~MeV). Including the reconstructed energy requirement, the electron-muon misidentification rate is assumed to be $<10^{-4}$ and is deemed a negligible background. Backgrounds due to cosmic ray interactions and the neutral current interaction $\nu_\mu e^- \rightarrow \nu_\mu e^-$ are considered insignificant as well. 

A 25\% systematic uncertainty on the background rate inside the signal region is used for this study. This number can be considered a conservative estimate and will ultimately depend on the understanding of detector/nuclear effects, charged/neutral kaon production uncertainties, and the $\textit{in-situ}$ background measurements available. Even with a conservative background uncertainty estimate, the eventual extracted sensitivity is limited by statistics.

\section{Sensitivity and discussion}
The figure of merit for this single bin, counting only experimental study is sensitivity to the LSND allowed region near $\Delta m^2\sim 1~\mathrm{eV}^2$ assuming a two neutrino oscillation hypothesis. The two neutrino oscillation probability is written as
\begin{equation}
P(\nu_\mu \rightarrow \nu_e)=\sin^2(2\theta_{\mu e})\sin^2\bigg(1.27\frac{\Delta m^2}{\mathrm{eV^2}}\frac{L}{\mathrm{km}}\frac{\mathrm{GeV}}{E}\bigg)~.
\end{equation}

\begin{figure}[h!]
\begin{center}
\begin{tabular}{c c}
\includegraphics[scale=.41]{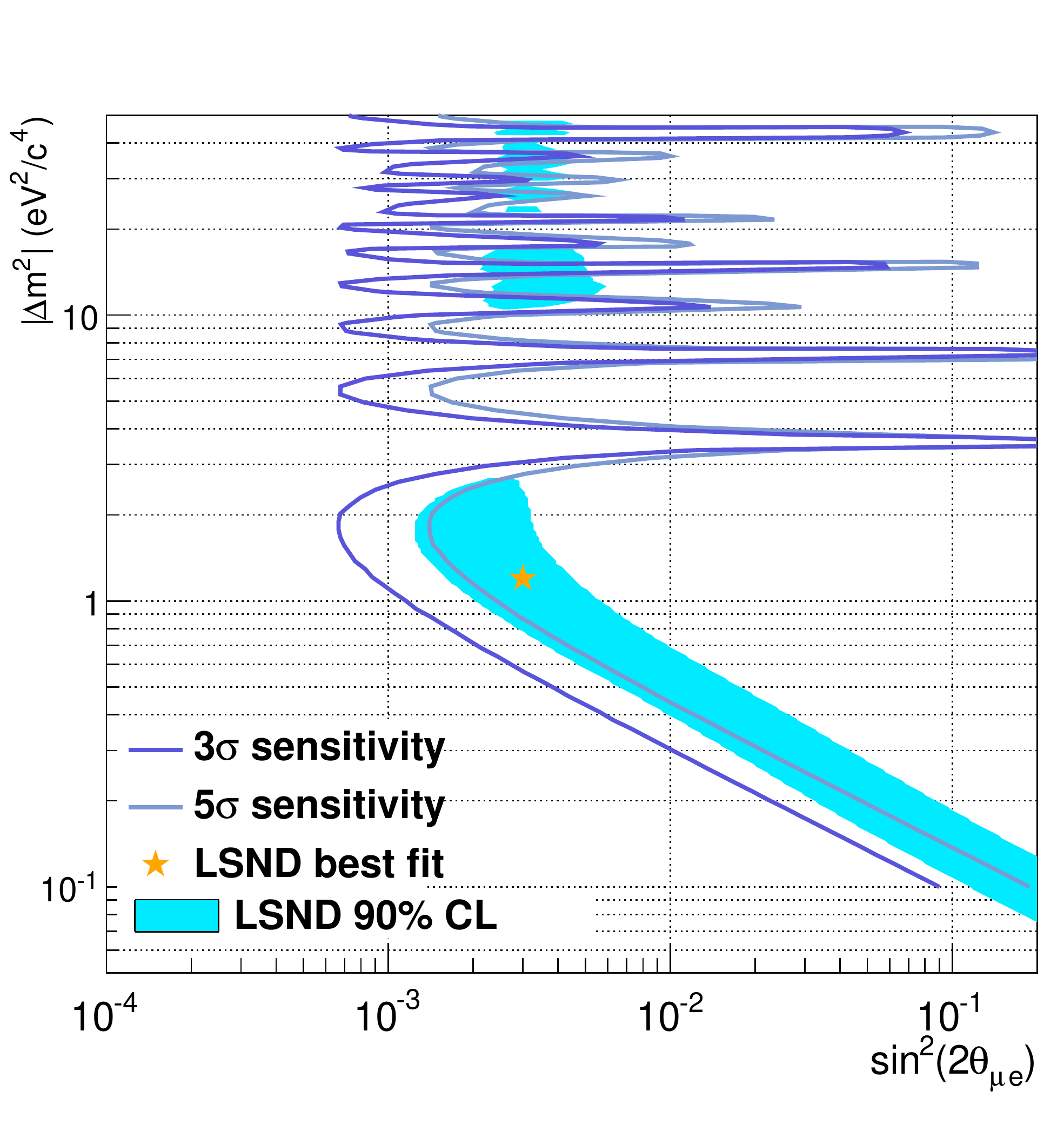}
\end{tabular}
\end{center}
\vspace{-.8cm}
\caption{The sensitivity to the LSND allowed region achievable with a kaon decay-at-rest source in combination with a large LArTPC. }
\label{fig:sensitivity}
\end{figure}

The constant baseline and neutrino energy can be inserted into the equation to form $P(\nu_\mu \rightarrow \nu_e)=\sin^2(2\theta_{\mu e})\sin^2(0.86\Delta m^2)$. The sensitivity curves, drawn in $[\sin^2(2\theta_{\mu e}),\Delta m^2]$ space and shown in Fig.~\ref{fig:sensitivity}, are based on calculations of fully frequentist confidence intervals using the profile log-likelihood method~\cite{rolke1,*rolke2}. ``Sensitivity'' is derived from the median upper limit that would be obtained by a set of experiments measuring background with no true signal~\cite{feldmancousins}. The uncertainty on the background, estimated at 25\%, is included as a nuisance parameter in the calculations. Baseline smearing effects due to detector length and spread in neutrino creation point are neglected.

The sensitivity curves show that a 2$\times 10^{21}$~POT exposure with Fermilab's 8~GeV booster in combination with a 2~kton LArTPC could either confirm or refute the controversial LSND result at 5$\sigma$ across most of the allowed region. In the case that a large $\sin^2(2\theta_{\mu e})$ is observed, such an experiment may also be able to provide sensitivity to the disappearance of $\nu_e$ [$\sin^2(2\theta_{e e})$] and/or $\nu_\mu$ [$\sin^2(2\theta_{\mu \mu})$] through the three and two body $K^{+}$ decay-at-rest, respectively. Given a localized neutrino source and an extended detector, observation of the ``oscillation wave'' as a function of $L/E$ is also possible.

One of the advantages of a low signal, low background experiment is that any event that enters the signal region can be carefully examined. Candidate signal interactions can be studied with the wide range of variables available for event characterization in a LArTPC, including a complete three dimensional picture of the neutrino-induced charged particle tracks~\cite{rubbia}. The search for a few $\nu_e$ candiate events among thousands of similar reconstructed energy $\nu_\mu$ events is made less daunting by this fact. The experimental concept is analogous to a dark matter search with only a handful of signal events expected among a large swath of easily distinguishable background events. 

There are a number of relatively uncertain assumptions used in this study that are worth pointing out: (1)~reconstructed neutrino energy resolution, in consideration of the nuclear physics affecting the initial interaction, the poorly understood intra-nuclear process leading to the eventual final nucleonic state, and the estimate for LArTPC proton reconstruction capability, (2)~background systematic uncertainty, (3)~kaon-induced neutrino flux, from decay-at-rest and decay-in-flight, and (4)~neutrino cross section. As discussed previously, the estimated uncertainties associated with each of these assumptions and, more importantly, their effect on sensitivity will be significantly reduced when the experiment turns on and begins taking data. Note that conservative estimates for both kaon production and neutrino cross section, each predicted with two different simulation packages, have been employed. Taking the larger, perhaps less conservative, of each of the two predictions would naively increase the expected signal (and background) by over 50\%. 

\section{Conclusion}
An experiment based on an intense kaon decay-at-rest source and a large LArTPC detector could provide a definitive probe of the LSND result and neutrino oscillations near $\Delta m^2\sim 1~\mathrm{eV}^2$. The source would also be an excellent opportunity for pion/muon decay-at-rest based oscillation searches (e.g.~\cite{oscsns}), a coherent neutrino-nucleus scattering program (e.g.~\cite{clear,coherentdaedalus}), and experiments to measure neutrino cross sections relevant for astrophysics (e.g.~\cite{nusns}), among other possibilities.

Although the detector mass and exposure used for this study (2~kton liquid argon detector with 2$\times 10^{21}$~POT from Fermilab's 8~GeV booster) can be considered optimistic, the idea of using the monoenergetic $\nu_\mu$ from charged kaon decay-at-rest for a $\nu_e$ appearance search is one that should be considered for current and future intense $\gtrsim$3~GeV proton sources around the world.

\begin{center}
{ {\bf Acknowledgments}}
\end{center}
The author wishes to thank P. Fisher, B. Fleming, T. Smidt, L. Winslow, and G. Zeller for useful suggestions; C. Andreopoulos, J. Sobczyk, and N. Mokhov for assistance with GENIE, NuWro, and MARS, respectively; G. Karagiorgi and M. Shaevitz for helpful comments and the oscillation sensitivity software framework; and J. Conrad for support and valuable insights.

\bibliography{kdarbib}
\end{document}